\documentclass{article}%
\usepackage{amssymb}
\usepackage{graphicx}
\usepackage{amsmath}%
\usepackage{amsfonts}
\begin{document}

\begin{center}
{\Huge Spinfluid Phase Transitions\bigskip}\bigskip

\bigskip Marcus S. Cohen

Department of Mathematical Sciences

New Mexico State University

marcus@ nmsu.edu
\end{center}

{\Large Abstract}

We start with the \textit{spinfluid: }a nearly-homogeneous, 8-spinor medium,
with small local $spin(4,%
%TCIMACRO{\U{2102} }%
%BeginExpansion
\mathbb{C}
%EndExpansion
)$ eddies and twists. As it expends, these seed a raft of intersecting
codimension $J$ \textit{singularities: }a \textit{spinfoam, }$\Sigma.$ As
$\Sigma$ expands, the energy trapped in each $(4-J)$ \ brane varies as
$\gamma^{J},$ where $\gamma(t)\equiv\frac{a(t)}{a_{\#}}$ \ is the
\textit{scale factor. }Summing on $J=(0,1,2,3,4)$ creates a quartic
\textit{dilation } \textit{potential } $V(n_{J};\gamma)\sim n_{J}\gamma^{J},$
with either 1 or 2 minima: \textit{preferred length and mass scales}. $\Sigma$
expands forever for $n_{4}<0,$ but recontracts for $n_{4}>0.$

To quantize $\Sigma,$ we take a canonical \textit{ensemble }$\widehat{\Sigma}$
of spinfoams, with \textit{mean }$J$-brane populations $N_{J}$ immersed in a
\textit{heat bath} of $N_{0}$ \textit{vacuum spinors, }whose microstates
vastly outnumber the matter states\textit{. }Its evolution is governed by a
\textit{free energy} $g(\mathbf{N};\gamma).$ This admits \textit{phase
transitions } at \textit{two }critical scales , $\gamma_{1}$ and $\gamma_{3}%
,$separated by a triple point\textit{, }$\gamma_{2}$. Their \textit{critical
droplets }correspond to the varieties of \textit{leptons }and \textit{hadrons}%
, respectively. We identify\ $\gamma_{1}$ as \textit{inflation, }$\gamma_{2}$
as \textit{decoupling, }$\gamma_{3}$ as \textit{baryogenesis; }and the heat
bath of \textit{vacuum spinors }as \textit{dark} \textit{energy. \ }

\bigskip\ \ 

\section{\bigskip The Spin World}

\bigskip

Could all the structures we see today evolve from a nearly-homogenous,
maximally-symmetric \textit{spinfluid}?\textit{\ }Could the length and mass
scales of particles, organisms, galaxies and the expanding universe emerge
from a conformally-invariant action?

We show here how conformal-symmetry breaking can occur, provided that the
initial state contains topological dislocations; \textit{phase singularities}
of \textit{different codimension, }$J=(0,1,2,3,4).$ We employ the same
8-spinor action that gave us the particles as singularities and the fields in
the regular regime surrounding them; but the results hold for the class of
models with projective singularities.

Does this mean that the fields and singularities of today's world must be
"programmed in" on some initial surface, as in holographic models

[Banks, Fischler]; [Hertog et.al.]?

This is like asking if exactly the right shapes of dust motes must be present
to nucleate the snow crystals in a snow storm. \textit{Any} set of initial
shapes,\textit{\ }followed by almost any microscopic fluctions in updrafts and
vapor pressure, will radiate an enormous variety of snowflakes near the
critical point for the phase transition.

We find two phase transitions in the spinfluid vacuum here, which we identify
as inflation and baryogenesis.

This is the 2nd paper in a 3-part series. In part \textrm{1) [M.C.6], }we
discovered that the families of elementary particles correspond to the classes
of codimension-$J$ singularities or \textit{caustics,}\textbf{\ }$\langle
p,q,r\rangle_{s}^{J}$ \ in a $spin(4,%
%TCIMACRO{\U{2102} }%
%BeginExpansion
\mathbb{C}
%EndExpansion
)$ phase flow, and their symmetry types to images in 3-mirror Kaleidoscopes
with dihedral angles $(\frac{\pi}{p},\frac{\pi}{q},\frac{\pi}{r}).$ Their rest
energies, $m_{s}=(\frac{s}{2})^{3},$ appeared in terms their
\textit{multiplicities}, $s:$ the number of reflections it takes to close a
cycle of \textit{null zigzags}. These calculated rest energies agreed with the
observed particle masses to within a few percent.The starting point there was the

{\LARGE Spin Principle}\textit{,}{\LARGE \ Pl. }\textit{\ Spinors, the
fundamental spin-1/2 repreesentations of spin isometries, are the primary
physical objects. Spacetime, geometry, gauge, and matter fields, along with
their interactions, emerge as spin tensors in }$4$\textit{\ left and }%
$4$\textit{\ right chirality spinors: }$8$\textit{\ in all, because it takes 8
spinors to make a natural 4 form.}

But spinors\textit{\ }are \textit{lightlike. }How can lightlike rays make a
\textit{massive} particle?

Mass is the ability of energy to stay in one place-i.e.for its lightlike
chiral components to weave a timelike worldtube, instead \ of zooming off at
the speed of light. The key is that \textit{counterpropagating }left and right
helicity spinors can make a standing wave, if their 3-momemta cancel, but
their energies add\textit{. }Counterpropagating chiral pairs with opposite
helicities but the same spins make\textit{\ massive} particles with spin, like
electrons, $e_{-}\in(l_{-}\oplus r_{-}),$ and positrons, $e^{+}\in(l^{+}\oplus
r^{+}).$ \textit{Copropagating} pairs with the same helicities make
\textit{massless} particles, like neutrinos, $\nu_{e}^{\circlearrowright}%
\in(l_{+}\oplus r_{-})$ and photons,$\ \gamma^{\circlearrowright}\in
(l_{+}\otimes r_{-}).$ A reaction like pair anhialation is just an exchange of
chiral partners in the 8- spinor basis:

\begin{center}
$e_{-}\otimes e_{+}=(l_{-}\oplus r_{-})\otimes(l^{+}\oplus r^{+})\rightarrow
l_{-}\otimes r^{+}\oplus r_{-}\otimes l^{+}=\gamma^{\circlearrowleft}%
\oplus\gamma^{\circlearrowright}.$
\end{center}

On a microscopic scale, each spinor or\textbf{\ }cospinor propagates along a
piecewise \textit{lightlike ray }segment: a "zig" outward- in the direction of
cosmic expansion, $\Delta T>0,$ or a "zag" inward, towards the "big bang",
$\Delta T<0:$

\begin{center}
$(\mathbf{\Delta}T,\Delta\mathbf{x)\in}$ $[r(1)+su(2)]_{diag}:\left\vert
\mathbf{\Delta x}\right\vert =\pm\Delta T=\Delta t,$
\end{center}

where $t$ is arctime: the arclength parameter along rays, and $T$ is cosmic
time: the logradius in $R_{4},$ or "imaginary time".

Quantum Field theory is statistical mechanics in imaginary time. The
statistical mechanics of the ensemble of null zigzags histories is greatly
simplified when the local system is immersed in a heat bath with a vastly
greater number of states: a stochastic background of \textit{vacuum spinors,}
with temperature $\beta^{-1}\thicksim$ $\hbar.$ Instead of summing over
creations and anhialations of intermediate particles, \textit{sum over the
microhistories of null zigzags on a lightlike lattice in "imaginery time", T,
connecting initial and final states.}

To compute these path integrals we start with a Lagrangian that is a\ natural
4 form, whose action is invarient under the Einstein group, \textbf{E}, of
spacetime (external) transformations: translations, rotatioms, boosts, and P
(space) and T (cosmic time) reversal- together with their spin space
(internal) representations. It takes \textbf{8} spinors - \textbf{4 }column
spinors, $\psi_{I}=\{l_{+},r_{+},l_{-},r_{-}\},$ and \textbf{4} provisionally
independent row spinors,$\psi^{I}=\{l^{+},r^{+},l^{-},r^{-}\}$, each of
conformal weight (dimension) $\frac{1}{2},$ to make a natural 4 form in
\textit{spin space}:\textit{\ }the 8-spinor position-velocity phase space
$(\psi\mathbf{,d}\psi).$ In complex coordinates on the cotangent bundle,

\begin{center}
$\mathbf{\Psi}=\bigcup_{I=1}^{4}(\psi^{I}+i\mathbf{d}\psi^{I}),(\psi
_{J}+i\mathbf{d}\psi_{J})\mathbf{\in}%
%TCIMACRO{\U{2102} }%
%BeginExpansion
\mathbb{C}
%EndExpansion
\mathbf{T}^{\ast}\psi,$\textit{\ }
\end{center}

where $\mathbf{d}\equiv e^{\alpha}\left(  x\right)  \partial_{\alpha}\left(
x\right)  $ is the generalized exterior differential operator.

The simplest natural 4 form, with only one kind of\ term and\textit{\ no
coupling constants, }is the $\mathbf{8}$-spinor factorization of the
Maurer-Cartan $4$ form; the \textbf{E-}invarient (scalar) measure on phase
space [M.C.1]:

\begin{center}
\bigskip$\mathcal{L}_{g}\equiv{\Huge \wedge}_{I,J=1}^{4}(\psi^{I}$
$\mathbf{d}\psi_{J})\equiv(\psi^{I}\mathbf{d}\psi_{J})^{\wedge4}.$
\end{center}

\bigskip Its action gives the\textit{\ volume in spin space:}

\begin{center}
$S_{g}=$ $\int(\psi^{I}\mathbf{d}\psi_{J})^{\wedge4}=\int r^{\pm}%
\mathbf{d}\ell_{\mp}\wedge\ell^{\pm}\mathbf{d}r_{\mp}\wedge\ell^{\mp
}\mathbf{d}r_{\pm}\wedge r^{\mp}\mathbf{d}\ell_{\pm}$
\end{center}

(sum over neutral sign combinations: i.e. whose $\pm$ signs $sum$ to 0). In
the \textit{stationery regime}, the cospinors are the Dirac $(P)$ conjugates
of the spinors: $\psi^{I}$ $\rightarrow\psi_{I}^{T}i\sigma_{2}.$The potential
energy, $V,$ is minimized when each spinor $\psi_{I}$ pairs with a
$P-$conjugate cospinor $\psi^{I}$. It is this \textit{attraction between
opposite-chirality spinors} that\textit{\ }stabilizes\textit{\ }%
particles\textit{. }\qquad\qquad\qquad\qquad\qquad\qquad\qquad\qquad

Spacetime enters as the parameter space for the action integral,
$S_{g}.\ \mathcal{S}_{g}$ is stationarized in either the PT symmetric
(gravitostrong) or PT antisymmetric (electroweak) case; in both, it gives the
covering number of the compactified internal group, $\ g=[U(l)xSU(2)]$, over a
compactified world-tube, or 4-brane, $B_{4}\backsim S_{l}\times S_{3}:$

\begin{center}
$S_{g}\rightarrow\frac{1}{2}$ $\int_{{\Large B}_{4}}Tr$ $[g^{{\Large -1}%
}dg]^{4}=16\pi^{3}$ $N$
\end{center}

\textit{\ }In the regular regime\textit{,} $S_{g}$ yielded the proper
effective actions for electroweak and gravitostrong fields, when all 8 spinor
fields ride on a nontrivial global background of \textit{vacuum spinors}: the
null spinors on compactified Minkowsky space, $M_{\#}\backsim S_{l}\times
S_{3}.$[M.C.1]. Accordingly, we expand each of the $\mathbf{8}$ spinor fields
here as a sum of a global, order-$k^{\frac{1}{2}}$ vacuum $\left(
\symbol{94}\right)  $ distribution, plus a local perturbation,
or\textit{envelope modulation}:

\begin{center}
$%
\begin{array}
[c]{cc}%
\ell^{\pm}\left(  x\right)  =k^{\frac{1}{2}}\lambda^{\pm}\left(  x\right)
+\xi^{\pm}\left(  x\right)  & r^{\pm}\left(  x\right)  =k^{\frac{1}{2}}%
\rho^{\pm}\left(  x\right)  +\eta^{\pm}\left(  x\right) \\
\ell_{\pm}\left(  x\right)  =k^{-\frac{1}{2}}\lambda_{\pm}\left(  x\right)
+\xi_{\pm}\left(  x\right)  & r_{\pm}\left(  x\right)  =k^{-\frac{1}{2}}%
\rho_{\pm}\left(  x\right)  +\eta_{\pm}\left(  x\right)  \text{.}%
\end{array}
$
\end{center}

In paper 1 [M.C.6], we classified\ the Lagrangian \textit{singularities,} or
\textit{spin caustics }$\gamma_{4-J}$ of codimension $J=(1,2,3,4):$
\textit{focal loci }where rays cross in the spacetime projection of the
\textbf{8}-spinor flow, and geometrical optics breaks down\ [M.C.2]\textit{.
}Caustics come in the discrete families, with the symmetry types in internal
space of images in 3-mirror Kaleidoscopes, with dihedral angles $(\frac{\pi
}{p},\frac{\pi}{q},\frac{\pi}{r});$ the \textit{Coxeter groups, of
multiplicity-}$s:$

\begin{center}
$\gamma_{s}\equiv$ $\langle p,q,r\rangle_{s}.$
\end{center}

The \textit{multiplicity, }$s,$ or \textit{Coxeter number, }is\textit{\ }the
minimal number of reflections it takes to close a zig-zag cycle of rays.

The surprising discovery in paper 1 was

\begin{center}
D1. \textit{Not only do the types of \textbf{8-}spinor caustics parallel the
families of leptons }$(J=1)$\textit{, mesons and photons }$(J=2)$\textit{, and
hadrons }$(J=3)$\textit{\ respectively, but their multiplicities }$s$
\textit{give their\textbf{\ correct masses}: }$m_{s}=(\frac{s}{2})^{3}$
(within a few percent).
\end{center}

For the electon, e$^{-}\equiv(l_{-}\oplus r_{-}),$ $\ s=2;$ the muon, $s=6,$
and the tauon, $s=30.$(see table 2, in paper1).

The key to the correspondence of caustics and particles lies in the
\textit{Dirac propagator}: the sum over \textit{null zigzags} connecting the
initial and final\textit{\ }states [Ord]. A\textit{\ null zig zag} is a double
reflection $l\rightarrow r\rightarrow l$ \ of \ (lightlike) spin rays
(characteristics of the massless Dirac operators, $D$ and $\overline{D})$ off
\textit{Clifford mirrors}: interference patterns with the 6
remaining\ \textit{vacuun spinor fields}. Penrose [Pen] called these
reflections \textit{mass scatterings}. Mass scatterings at the boundary are
what keep null zigzags confined to a world tube in cosmic time, $T$- i.e
\ what give a bispinor wavefunction \textit{mass}.

The \textit{statisticalcal mechanics} of matter envelopes in "imaginary time",
$T,$ is\textit{\ }the quantum field theory of particles in "real time", $t.$
Since the vacuum spinors- the \textit{dark matter and energy,} make up about
90\% \ of the total energy of the universe, the \textit{stochastic vacuum
}acts like a \textit{heat bath }for the matter spinors, which ride on the
vacuum like waves on the surface of the ocean.We derive a free energy, or
\textit{dilation potential} $g(N;\gamma),$for the vacuum seeded homogeneosly
with $N_{J}$ topological detects of codimension $J=(1,2,3,4)$ which has either
1 or 2 minima: prefered length and mass scales. This dilation potential will
act as an effective \textit{Higgs field}, whose hills and valleys are sculpted
by the\textit{\ nonlocal \ effects }of all the topological defects
in\bigskip\ compactifieid spacetime: the "distant masses" of Mach's principle.

\bigskip

\section{Topological Trapping of Currents on Dual Branes}

\bigskip

The flux integral,%
\[
\int_{\gamma_{2}}K_{or}e^{\theta}\wedge e^{\varphi}\equiv\int_{\gamma_{2}}\ast
K=4\pi N\text{,}%
\]

seems to quantize \textit{electric field, }$K_{0r}$, over \textit{dual}
(transverse)\textit{\ }surfaces, $S_{\theta\varphi},$ spanned by the surface
element $e^{\theta}\wedge e^{\varphi}.$

Now the magnetic field, \ $K_{\theta\varphi},$ is quantized over spatial 2
surfaces by deRham cohomology. Why should the \textit{dual field} be quantized?

Every contribution to $S_{g}$ must be a Clifford $(C)$ scalar, built of all 8
spinors. The matter bispinors make the $C-$ valued $J$- form current,
$\omega^{J}.$ This must be multiplied by some $(4-J$ $)$ form, $\Omega^{4-J} $
, that is \textit{both }$C$\textit{\ and Hodge dual} to it to get a $C-$scalar
$(\sigma_{0})$ valued 4 form- the only kind that can be invariently
integrated. Where does this dual come from?

The key is quantization of the topological action\textit{\ }[M.C.2],

\begin{center}
\bigskip$S_{g}=\int_{\mathbb{M}_{\#}}\Omega^{\wedge4}=16\pi^{3}N.$
\end{center}

Table \textbf{I (}below) shows how the vacuum spinors on $\mathbb{S}_{1}%
\times\mathbb{S}_{3}$ \textit{automatically} provide \textit{\ }$(4-J$ $)$
surface elements \textit{dual }to any $J-$ bispinor matter current; the ones
needed to quantize the normal flux.%

\[%
\begin{array}
[c]{c}%
\text{\textbf{Table I}}\\
\text{\textbf{Vacuum Spin Forms:} Exterior Products }\hat{\Omega}^{J}\text{ of
the \emph{vacuum spin connections} on }\mathbb{S}_{1}\times\mathbb{S}%
_{3}\left(  a_{\#}\right)  ,\\
\text{with compactification radius }a=\gamma a_{\#}\text{, and Minkowsky
metric.The upper sign on }\sigma_{j}\text{ is the }L\text{-chirality part;}\\
\text{the lower sign, the }R\text{. The upper and lower signs on }\sigma
_{0}\text{ apply to the analytic and conjugate-analytic representations of
time translations.}%
\end{array}
\]

\begin{center}
$%
\begin{array}
[c]{c}%
\hat{\Omega}=\pm\left(  \frac{ik}{2a_{\#}}\right)  \sigma_{\alpha}e^{\alpha}\\
\hat{\Omega}^{2}=\left(  \frac{ik}{2a_{\#}}\right)  ^{2}\sigma_{\ell}\left[
\epsilon_{jk}^{\;\ell}e^{j}\wedge e^{k}\pm e^{0}\wedge e^{\ell}\right] \\
\hat{\Omega}^{3}=\pm\left(  \frac{ik}{2a_{\#}}\right)  ^{3}\sigma_{\ell
}\epsilon_{jk}^{\;\ell}e^{j}\wedge e^{k}\wedge e^{0}\pm i\epsilon_{jk\ell
}\sigma_{0}e^{j}\wedge e^{k}\wedge e^{\ell}\\
\hat{\Omega}^{4}=\left(  \frac{ik}{2a_{\#}}\right)  ^{4}\sigma_{0}\left[
\epsilon_{\alpha\beta\gamma\delta}e^{\alpha}\wedge e^{\beta}\wedge e^{\gamma
}\wedge e^{\delta}\right]  =\frac{3}{2}\left(  \frac{k^{4}}{a_{\#}^{4}%
}\right)  \mathbf{d}^{4}v
\end{array}
$

The vacuum\ spin forms, $\hat{\Omega}^{\left(  4-J\right)  }$, make Clifford
line, surface, and volume elements

dual \ to the $J$ pairs of matter spinors and $PT$-conjugate differentials
$\left(  \psi^{I}\mathbf{d}\psi_{I}\right)  ^{J}$:

the ones they multiply to give the \emph{Clifford scalar }volume element
$\sigma_{0}e^{0}\wedge e^{1}\wedge e^{2}\wedge e^{3}\mathbf{d}^{4}v$.

\bigskip
\end{center}

For stationery solutions, our 8-spinor Lagrangian reduces to the Maurer-Cartan
4 form- the volume form in $U(1)\times SU(2),.$whose action is topologically
quantized over $\mathbb{M}_{\#}\sim\mathbb{S}_{1}\times\mathbb{S}_{3}.$

Alternatively, as Witten has pointed out, the \emph{Weiss-Zummo} 4 form may be
quantized over the boundary, $\gamma_{4}\sim\partial B_{5}$ of the 5-manifold,
$B_{5}\sim\mathbb{C}\times\mathbb{S}_{3}$, obtained by complexifying
\emph{time:} $z^{0}\equiv t+iT\in\mathbb{C}$ --- and then imposing periodic
(or matched asymptotic) boundary conditions on the inertial and final
hypersurfaces [Witten 1,2]:%
\[
\int_{\partial B_{5}}\zeta\left(  \mathbf{d}\zeta\right)  ^{4}=32\pi
^{4}N\text{.}%
\]

The intensity, $k$, of each chiral pair of vacuum spinors must scale as
$k\sim$ $\gamma^{-1}$ for the action of the $4$ global pairs, quantized over volume

$V\left(  T\right)  =$ $\gamma^{4}V\left(  0\right)  $, to remain constant.
Then the integral of the $J$ \emph{matter-spinor} pairs, $\tilde{\Omega}^{J}$,
over their singular loci, $B_{J}$ remains constant, while the dual vacuum spin
forms $\hat{\Omega}^{4-J}$, contribute a factor of $k^{4-J}\sim\gamma^{J-4}$.
After multiplying by the uniformly expanded 4-volume element, $\left(
\mathbf{d}V\right)  ^{4}=\gamma^{4}\left[  \mathbf{d}v\right]  ^{4}$, this
gives a net factor of $\gamma^{J}$ in the action contributed by the $D^{J}$ stratum:

\begin{center}
\bigskip

$%
\begin{array}
[c]{cc}%
J=0\text{ (vacuum spinors);} & \int_{\widehat{\mathbb{M}}}\widehat{\Omega}%
^{4}=16\pi^{3}%
\begin{array}
[c]{c}%
\\
n_{0}\text{;}\\
\end{array}
\\
J=1\text{ (leptons):} &
\begin{array}
[c]{c}%
\underset{D^{1}}{\sum}\int_{S_{3}\times\gamma_{1}}\widehat{\Omega}^{3}%
\wedge\widetilde{\Omega}^{1}\\
-i8\pi^{2}\gamma\underset{D^{1}}{\sum}i2\pi m\equiv16\pi^{3}n_{1}\gamma
\end{array}
\\
& \\
& \\
J=2\text{ (photons; mesons):} &
\begin{array}
[c]{c}%
\underset{D^{2}}{\sum}\int_{S_{2}\times\gamma_{2}}\widehat{\Omega}^{2}%
\wedge\widetilde{\Omega}^{2}\\
=4\pi\gamma^{2}\underset{D^{2}}{\sum}4\pi q\equiv16\pi^{3}n_{2}\gamma^{2}%
\end{array}
\\
& \\
J=3\text{ (hadrons):} &
\begin{array}
[c]{c}%
\underset{D^{3}}{\sum}\int_{S_{1}\times\gamma_{3}}\widehat{\Omega}^{1}%
\wedge\widetilde{\Omega}^{3}\\
=i2\pi\gamma^{3}\underset{D^{3}}{\sum}i8\pi^{2}b\equiv16\pi^{3}n_{3}\gamma^{3}%
\end{array}
\\
& \\
J=4\text{ (spin }\left(  4,\mathbb{C}\right)  \text{ vortices; atoms):} &
\underset{D^{4}}{\sum}\int_{\gamma_{4}}\widetilde{\Omega}^{4}=\underset{D^{4}%
}{\sum}16\pi^{3}n_{4}\gamma^{4}%
\end{array}
$
\end{center}

The sum of terms (\ref{17}) gives the net action expressed in powers of
$\gamma$(\ref{17}) :%
\begin{equation}%
\begin{array}
[c]{c}%
\widehat{S}_{g}\left(  \gamma\right)  =\int_{B_{0}}\hat{\Omega}^{4}+\gamma
\int_{B_{1}}\hat{\Omega}^{3}\wedge\left(  \psi^{I}\mathbf{d}\psi_{I}\right)
+\gamma^{2}\int_{B_{2}}\hat{\Omega}^{2}\wedge\left(  \psi^{I}\mathbf{d}%
\psi_{I}\right)  ^{2}\\
+\gamma^{3}\int_{B_{3}}\widehat{\Omega}\wedge\left(  \psi^{I}\mathbf{d}%
\psi_{I}\right)  ^{3}+\gamma^{4}\int_{B_{4}}\left(  \psi^{I}\mathbf{d}\psi
_{I}\right)  ^{4}\\
=16\pi^{3}\left[  n_{0}+n_{1}\gamma+n_{2}\gamma^{2}+n_{3}\gamma^{3}%
+n_{4}\gamma^{4}\right]  \text{.}%
\end{array}
\label{18}%
\end{equation}

The Euclidean action integral (\ref{18}) over $\left(  T,\mathbf{x}\right)  $
serves as a Lagrangian%
\begin{equation}
\left(  \gamma,\overset{\cdot}{\gamma}\right)  =T\left(  \dot{\gamma}\right)
-V\left(  \gamma\right)  \text{,}\label{11a}%
\end{equation}
governing the evolution of the scale factor $\gamma\left(  t\right)  $ in
\emph{Minkowsky} time: the "many-fingered" time parameter common to all
particle trajectories and piece-wise null rays (characteristics), $\left\vert
\bigtriangleup T\right\vert =\left\vert \bigtriangleup\mathbf{x}\right\vert
=\bigtriangleup t$.

The \textit{potential energy},
\[
V\left(  \gamma\right)  =-\left[  n_{0}+n_{1}\gamma+n_{2}\gamma^{2}%
+n_{3}\gamma^{3}+n_{4}\gamma^{4}\right]  \text{,}%
\]
contains the standard pressure terms. The radiation pressure, $p_{r}$, varies
as $\gamma^{-4}$. Its $4$-volume integral is contained in the constant
($n_{0}$) term: the total energy stored in the background radiation: cosmic
neutrinos and photons. The matter pressure\emph{,}$p_{m}$, varies as
$\gamma^{-3}$, and contributes to the linear ($n_{1}$) term.

The \textit{kinetic energy}\emph{,} $T\left(  \dot{\gamma}\right)  $, of a
Friedmann 3-brane expanding in the ambient space comes from the balance
between the analytic ($+$) and conjugate analytic ($-$) spin waves, whose
interference pattern strikes off this 3-brane physically as the level surface
of the $U\left(  1\right)  $ phase, $\theta^{0}\equiv\theta_{+}^{0}+\theta
_{-}^{0}=0;$ the \textit{bulk-neutrality} condition for the background $u(1)$
flux, as measured in a coexpanding frame. Here%
\[
\zeta_{\pm}^{0}\equiv\left(  \theta_{\pm}^{0}+i\varphi_{\pm}^{0}\right)
\left(  t\pm iT\right)
\]
are the complex phase factors of the analytic ($+$) and conjugate-analytic
($-$) spinors respectively. They obey the Cauchy-Riemann conditions

\begin{center}
$\partial_{t}\varphi_{\pm}^{0}=\mp\partial_{T}\theta_{\pm}^{0}$;\qquad
$\partial_{T}\varphi_{\pm}^{0}=\pm\partial_{t}\theta_{\pm}^{0} $.
\end{center}

Bulk-neutrality, together with analyticity, gives a net dilation exponent of%
\[
\varphi\equiv\frac{1}{2}\left[  \varphi_{+}^{0}-\varphi_{-}^{0}\right]
=\varphi_{+}^{0}=-\varphi_{-}^{0}%
\]
for our Friedman 3-brane solution, $S_{3}(t).$ It is governed by the effective
Lagrangian%
\[
L\left(  \varphi,\overset{\cdot}{\varphi}\right)  =\overset{\cdot}{\varphi
}+\ln P\left(  \varphi\right)  ,
\]
where%
\[
P\left(  \varphi\right)  =n_{0}e^{-2\left(  \varphi\right)  }+n_{1}e^{-\left(
\varphi\right)  }+n_{2}+n_{3}e^{\left(  \varphi\right)  }+n_{4}e^{2\left(
\varphi\right)  }\text{.}%
\]
Varying $L$ with respect to $\varphi$, we obtain the differential equation
governing the Minkowsky-time evolution of our expanding shell, $\mathbb{\hat
{M}}=\left(  T\left(  t\right)  ,S_{3}\left(  t\right)  \right)  $:%
\[%
\begin{array}
[c]{c}%
\overset{\cdot\cdot}{\varphi}=-\partial_{\varphi}\ln P\left(  \mathbf{n}%
;\varphi\right)  \text{;}\\
P\left(  \mathbf{n};\varphi\right)  =n_{0}e^{-2\varphi}+n_{1}e^{-\varphi
}+n_{2}+n_{3}e^{\varphi}+n_{4}e^{2\varphi}\text{,}%
\end{array}
\]
where $\left(  \varphi_{I}+\varphi^{I}\right)  \equiv\varphi\left(
\mathbf{n};t\right)  $ is the net dilation exponent. Here $\mathbf{n}%
\equiv\left(  n_{0},n_{1},n_{2},n_{3},n_{4}\right)  $, the \emph{population
vector,} is the number of cells in each dimension. We consider $\mathbf{n}$
constant first. Below, we assume that $\mathbf{n}\left(  \gamma\right)  $
adjusts more quickly to $\gamma$ than $\gamma$ can change in $t$ --- i.e. that
$\mathbf{n}$ is "slavad" to $\gamma\left(  t\right)  $. A more detailed model
could include the\emph{\ connectivity} of the cells, say written as an
incidence matrix. Like a raft of soap bubbles, the connectivity of the complex
could change with $t$ as tubes $\gamma_{4}$, $\gamma_{3}$, $\gamma_{2}$,
$\gamma_{1}$, and $\gamma_{0}$ divide and reconnect. But there is a
constraint: the alternating sum of the coefficients, or the \emph{index} of
the $Spin^{c}$-$4$ complex,%
\[
\mathcal{I}=16\pi^{3}\left[  n_{0}-n_{1}+n_{2}-n_{3}+n_{4}\right]  \text{,}%
\]
is a homotopy invariant, that must be conserved over $t$ evolution..

The dilation rate $\varphi$ oscillates in the effective potential well:%
\[
U\left(  \mathbf{n};\varphi\right)  \equiv-\ln\left[  n_{0}e^{-2\varphi}%
+n_{1}e^{-\varphi}+n_{2}+n_{3}e^{\varphi}+n_{4}e^{2\varphi}\right]  \equiv-\ln
P\left(  \mathbf{n};\varphi\right)  \text{.}%
\]
The potential gradient, $\partial_{\gamma}U$, will pull $\varphi\left(
t\right)  $ downhill in $U\left(  \gamma\right)  $, to end up in a local
minimum; or in a global minimum, if there is enough kinetic energy
$\overset{\cdot}{\varphi}\left(  t\right)  $ to "coast" over any intervening
uphill sections.

For example, for cell population vector%
\[
\mathbf{n}=\left(  e^{-4},-e^{-1},1,e^{-1},-e^{4}\right)  \text{,}%
\]
the system has a stable equilibrium at $\varphi_{s}\simeq-2.35$, $U_{s}%
\simeq-0.135$, and an \emph{unstable} equilibrium at $\varphi_{u}\simeq2.35$,
$Uu=1.518$ (Figure 1).

The first equilibrium, $\varphi_{s}$, represents a classical \emph{soliton}; a
standing wave bound state inside its own potential well. Physically, it is a
3-sphere $\mathbb{S}_{3}\left(  a_{s}\right)  $, covered $n_{0}$ times by
$SU\left(  2\right)  _{L}\times SU\left(  2\right)  _{R}$, for which the
attractive force between hadrons ($n_{3}$) is stably balanced by the
quantum-mechanical preference of leptons ($n_{1}$) for \emph{delocalization}.
This topologically-nontrivial field configuration cannot be captured by ad-hoc
cutting and pasting of general relativity and quantum field theory; it lives
within the domain of the unified theory, which gives rise to them both.

We expect oscillations about the local minimum $\varphi_{s}$. But if $\varphi$
exceeds local maximum $\varphi_{u}$, the state will escape its basin of
attraction and $\varphi$ will \emph{inflate} rapidly to the next branch, which
is attractive for $n_{3}>0$. For $n_{4}<0$, this branch is unstable, and the
radius expands forever. For $n_{4}\geq0$, however, the state gets "caught" in
the basis of a second stable equilibrium.

We show below that the statistical \emph{ensemble} of $Spin^{c}$-$4$ complexes
exhibits an \emph{inflationary phase transition} at the \emph{critical }radius
$a_{c}=a_{\#}$, \emph{before} the classical system becomes unstable. As $a$
increases further, there is a "critical cascade" of $J=$ $(1,2,3,and4)$ pairs
of matter spinors condensing on the codimension-$J$ singular loci, $D^{J}$.

\section{Pattern Nucleation Near the Critical Points}

\bigskip

Quantum mechanics is statistical mechanics in "imaginary time," i.e.
\emph{cosmic time:} the logradius $T=a_{\#}\ln\gamma$ of our spatial hypersurface.

$T$ combines with Minkowsky time, $t$, to make \emph{complex time, }
\[
z^{0}\equiv t+iT\text{.}%
\]

They are related by the "Wick rotation", $W$ ; analytic continuation to
imaginery time, $T$:

\begin{center}
$\quad%
\begin{array}
[c]{c}%
W:t\longrightarrow T\Rightarrow e^{-\beta E_{S}}\longrightarrow e^{i\hbar
^{-1}S_{P\text{ }}}\text{;}\\
E=T\left(  \dot{\gamma}\right)  +V\left(  \gamma\right)  \rightarrow-T\left(
\dot{\gamma}\right)  +V\left(  \gamma\right)  \equiv-S\left(  t,\mathbf{x}%
\right)  \text{.}%
\end{array}
$
\end{center}

But wait! This mathematical "sleight of hand", $W,$\ obscures the
\textit{underlying physics}:

\begin{center}
\emph{the interaction of every local system with an invisible heat bath of
\textbf{vacuum energy }at}

\emph{\ "temperature" }$\beta^{-1}=$\emph{\
%TCIMACRO{\U{127} }%
%BeginExpansion
\protect\rule{0.1in}{0.1in}
%EndExpansion
, that contains vastly many more microstates than any local system. }
\end{center}

This enormously simplifies the sum over microstates, because we need only
count the microstates $\delta\hat{\Omega}$ gained or lost by the
\textit{vacuum }in\textit{\ }exchanging a 4-dimensional energy increment,
$\delta\hat{E},$ with the local system.

By definition,%
\[
\beta\equiv\frac{\delta\ln\hat{\Omega}}{\delta\hat{E}}\Longrightarrow
\frac{\delta\hat{\Omega}}{\delta\hat{E}}=\beta\hat{\Omega},
\]
where $\delta\hat{E}$ is the incremental change in the energy of the
\emph{heat bath} need to create $\delta\hat{\Omega}$ new vacuum states. In the
continuum approximation,%
\[%
\begin{array}
[c]{c}%
\frac{\delta\hat{\Omega}}{\delta\hat{E}}=-\beta\hat{\Omega}\Longrightarrow
\hat{\Omega}\left(  \hat{E}_{0}+\delta\hat{E}\right)  =\hat{\Omega}\left(
\hat{E}_{0}\right)  e^{\beta\delta\hat{E}}\\
=\hat{\Omega}_{0}e^{-\beta\delta E_{S}}\simeq\hat{\Omega}\left(  \hat{E}%
_{0}\right)  \left(  1+\beta\delta\hat{E}\right)  \text{,}%
\end{array}
\]
since $\delta E_{s}=-\delta\hat{E}$.

However, states really come in integral units:

\begin{center}
$\delta\hat{\Omega}=1=\beta\delta\hat{E}\Longrightarrow\delta\hat{E}%
=\beta^{-1}=\hbar\rightarrow\delta S_{P}$
\end{center}

The quantum of action is the Euclidean energy \emph{needed to create one more
vacuum state;} because 4-energy is\emph{\ }\textit{action }under analytic
continuation to Minkowsky time,%
\[
W:\beta\delta\hat{E}\longrightarrow\hbar^{-1}\delta S_{P}\text{.}%
\]

Action is quantized because vacuum states are discrete; as are the microscopic
states of any localized system, say, $\Psi\left(  t\right)  \equiv\Psi\left(
T,\mathbf{x};t\right)  $: a \emph{microhisory} of the wave-function on the
lattice, $\left(  T,\mathbf{x}\right)  \in N$, as it evolves in arctime, $t$.
Wick rotation maps the energy $E_{C}$ of \ a 4-dimensional state
$\varphi\left(  T,X\right)  $ to the action $S\left(  \varphi\right)  $ for a
\emph{path: }a \emph{microhistory}, $\psi\left(  t\right)  $, parametrized by
Minkowsky time, $t$.

Now the macroscopicworld seems to behave like the average over the
\emph{ensemble}, $\sum_{C}$, of possible microstates compatible with the
macroscopic state. If there are competing states- in this case macroscopic
histories---it's not just the \emph{energy} $E$ of each state, but also its
\emph{entropy}%
\[
S\equiv k\ln\Omega\text{,}%
\]
the log of the number $\Omega$ of microscopic realizations in its ensemble,
that determines which macrohistory it will choose.

Energy and entropy combine in the \emph{Gibbs free energy}%
\[
G\equiv E+pV-TS\text{,}%
\]
which is minimized at equilibria: $dG=0.$ It is the \emph{entropy} term, $-TS
$, that distinguishes the statistical mechanics of an ensemble from the
classical mechanics of any of its members. States flow downhill in $G$;
$\Delta G\leq0,$ in any spontaneous process, because $\bigtriangleup
S=k\bigtriangleup\ln\Omega>0$, because a given energy increment transferred
from system to \emph{vacuum} may fill many more possible vacuum states than
the number of states lost to the system.

For a massive bispinor particle, the microhistories are \emph{null zig-zag}
paths, whose \emph{edges} are segments of the light-like world lines of $L$-
and $R$-chirality spinors, and whose \emph{vertices} are mass-scatterings with
the vacuum spinors.These edges $\gamma_{1}$ bound surfaces $\gamma_{2} $,
which bound volumes $\gamma_{3}$; these bound world tubes, $\gamma_{4}$. We
call the union of all of these simplices a \emph{Spin}$^{C}\emph{-4}$
\emph{complex.} Each complex can be inscribed on a \emph{null lattice}, $N$,
with null edges $\gamma_{1}$. The particle \emph{propagator} is the sum over
all null zig-zag paths connecting the initial and final vertices.

The count of possible histories of particles and interactions thus behaves as
if each were restricted to a piecewise-null \emph{lattice} $\left\vert
\frac{\bigtriangleup\mathbf{x}}{\bigtriangleup T}\right\vert =1$, with each
cycle supporting integral holonomy of the $u\left(  1\right)  \oplus su\left(
2\right)  $ phase in its spacetime projection,%
\[
\Pi:z\left(  t\right)  \longrightarrow\gamma_{1}\left(  t\right)
\equiv\left[  T\left(  t\right)  ,\mathbf{x}\left(  t\right)  \right]
\text{.}%
\]%
\[%
\begin{array}
[c]{c}%
\int_{\gamma_{1}}P\equiv\int_{\gamma_{1}}P_{\alpha}e^{\alpha}=\int_{\gamma
_{1}}Edt-\int\mathbf{P}\cdot\mathbf{dx}\\
=\int_{t_{1}}^{t_{2}}\left(  \overset{\cdot}{T}\partial_{T}\zeta
-\overset{\cdot}{\mathbf{x}}\partial_{\mathbf{x}}\zeta\right)  \mathbf{d}%
t\equiv\int_{\gamma_{1}}\mathbf{d}\zeta\equiv\bigtriangleup\zeta=n=2\pi
n\hbar\\
\left[  \hbar\pi\left(  n+\frac{1}{2}\right)  -\hbar\pi\left(  m+\frac{1}%
{2}\right)  \right]  \sigma_{0}\hbar\pi N\text{;}\\
\text{with }N\equiv n+m;\text{ an integer.}%
\end{array}
\]
This\textit{\ Integral Holonomy condition} enables wavefunctions to be
\emph{single-valued} on the null lattice, $N$.

Now each macroscopic history, $C$ admits an ensemble of microhistories, $s$,
with different cell \emph{counts,} $n_{J}^{s}$. We have assumed that
\emph{all} the microcells of codimension $J$ contribute \emph{equally} to the
potential energy, $V^{s}\left(  \gamma\right)  =-\left[  n_{J}^{s}\gamma
^{J}\right]  $, of each state.

Each $(4-J$ $)$ cell $B_{4-J}$ with energy $\epsilon_{J}$ contributes a
probability weight of $e^{-\beta\epsilon_{J}}$ to the ensemble average. A
complex, $s$, containing $n_{J}^{s}$ $J$ cells has probability weight
$e^{-\beta n_{J}^{s}\epsilon_{s}^{J}}.$

We assume \emph{equipartition}: $T^{s}=V^{s}$, so%
\[
\epsilon^{s}\left(  \gamma\right)  =-2n_{J}^{s}\gamma^{J}\text{.}%
\]

Summing on $J=\left(  0,...,4\right)  $, and over all the microhistories $s$
in the ensemble, we obtain the \emph{partition function},%
\[
Z\equiv%
%TCIMACRO{\dsum \limits_{s}}%
%BeginExpansion
{\displaystyle\sum\limits_{s}}
%EndExpansion
e^{-\beta\left(  n_{J}^{s}\epsilon_{\mathbf{s}}^{J}\right)  }\text{.}%
\]
Since the energy is extensive, the partition function factors:%
\[
Z=\frac{\zeta_{0}^{N_{0}}\zeta_{1}^{N_{1}}...\zeta_{4}^{N_{4}}}{N_{0}%
!N_{1}!...N_{4}!}\text{, \ where }\zeta_{J}\equiv%
%TCIMACRO{\dsum \limits_{s}}%
%BeginExpansion
{\displaystyle\sum\limits_{s}}
%EndExpansion
e^{-\beta\epsilon_{\mathbf{s}}^{J}}\text{.}%
\]
Here%
\[
N_{J}\equiv\bar{n}_{J}=\beta^{-1}\frac{\partial\ln Z}{\partial n_{J}}%
\]
is the \emph{mean} (ensemble-average) population of the $J$th stratum.

The calculation now goes just like the one for reacting chemical species in
equilibrium, except that conservation of atoms is replaced by conservation of
the \emph{Index,} $\mathcal{I}=\left(  -1\right)  ^{J}N_{J}$, as $\gamma$
grows and the simplices "react" to form a new minimal spin complex:%
\[
\sum\left(  -1\right)  ^{J}\mu_{J}=0\text{,}%
\]
where%
\[
\mu_{J}\equiv\frac{\partial G}{\partial N_{J}}=-\beta^{-1}\ln\left[
\frac{\zeta^{J}}{N_{J}}\right]
\]
is the \emph{chemical potential} for the $J$th component: again, the capital
letters indicate ensemble averages:%
\[
X\equiv\bar{x}=\beta^{-1}\frac{\partial\ln Z}{\partial x}\text{.}%
\]

At equilibrium%
\[
\zeta_{0}\zeta_{1}^{-1}\zeta_{2}\zeta_{3}^{-1}\zeta_{4}=K\text{,}%
\]
the equilibrium constant.

For example, suppose that the mean populations of only two strata, say $J$ and
$K$, were allowed to vary. Then
\[%
\begin{array}
[c]{c}%
\frac{\bigtriangleup N_{J}}{\bigtriangleup N_{K}}=-\gamma^{J-K}\text{;}\\
\text{e.g.\quad}\frac{\bigtriangleup N_{4}}{\bigtriangleup N_{0}}=-\gamma
^{4}\text{.}%
\end{array}
\]
As $\gamma$ increases the population shifts from the $J=0$ stratum, through
the $J=2$ and $3$ strata, to the $J=4$ stratum.

For any given temperature and pressure, it is the \emph{Gibbs free energy
}that is minimized at equilibrium:%
\[%
\begin{array}
[c]{c}%
G\left(  N_{J},\gamma\right)  \equiv E+PV-TS\text{;}\\
\bigtriangleup G\sim\bigtriangleup E+\bigtriangleup\left(  PV\right)
-T\bigtriangleup S=0\text{.}%
\end{array}
\]

So far, we have only accounted for the \emph{entropy }contribution,
$\bigtriangleup S$, from changes in the cell populations, $N_{j}$. We must add
the contribution of cosmic expansion, through the potential energy $V\left(
\gamma\right)  .$ \emph{Equipartition} mandates an equal contribution from the
kinetic energy,%
\[
E=T+V=2V=-2\sum N_{J}\gamma^{J}\text{,}%
\]
giving%
\[
\bigtriangleup G\sim\left(  \frac{\partial G}{\partial N_{J}}\right)
\bigtriangleup N_{J}+\left(  \frac{\partial G}{\partial\gamma}\right)
_{N}\bigtriangleup\gamma=\sum\mu^{J}\bigtriangleup N_{J}+\int\frac{\partial
G}{\partial\varphi}\mathbf{d}\varphi\text{.}%
\]

The effect of the integral is to create a \emph{phase transition} in the
statistical ensemble. The phase point $\left(  \varphi,G\left(  \varphi
\right)  \right)  $ moves past the local minimum $\left(  \varphi_{s}%
.G_{s}\right)  $ on the stable branch, but never reaches local maximum,
$\left(  \varphi_{u}.G_{u}\right)  $, where the classical system (Figure 1)
would have lost stability. Instead, when it reaches the \emph{critical point}
$\left(  \varphi_{c}.G_{c}\right)  $ , it jumps into (or "tunnels" through to)
the point $\left(  \varphi_{d}.G_{c}\right)  $, where a further increase in
$\varphi$ will again cause the integral to \emph{decrease.} The phase
transition occurs where the areas between the Maxwell line $U=U_{c}$ and the
curve $U\left(  \gamma\right)  $ on the left and right sides are equal.
Moreover,the fastest growing scale for fluctuations is the critical scale,
$\gamma_{c},$ for the\textit{\ inflationary phase transition.}

\bigskip

\subsection{Critical Pattern Growth and Selection}

\bigskip

The stochastic dynamics of cell populations suggests the following picture of
pattern formation with increasing scale factor.

\begin{enumerate}
\item For $\gamma\leq\gamma_{C}$ the fundamental lightlike modes; the vacuum
spinors, dominate the ensemble average. They carry the lowest energy,
$N_{0}\gamma^{4},$ and thus the highest weight,
\[
N_{0}\sim e^{-\beta\epsilon_{0}}.
\]
The fastest-growing wavelength for fluctuations \emph{remains the same} as the
doman inflates, causing the condensation of critical droplets. As $\gamma$
increases, the $N_{1}$term comes into play, stabilising the first basin of
attraction. Leptons bound in this basin stack up in higher and higher energy
states. The leptonic phase loses stability at the first \emph{critical point,
}$\gamma_{c},$ and inflation occurs. .

\item It is here that the entropy term $-TS$ in $G$ has a decisive effect, by
catalyzing the formation of a Bose condensate. Unlike Fermions, an unlimited
number of Bosons may be put into a state. This tends to raise the entropy. The
entropy contribution to the Gibbs free energy varies as the \emph{cube} of the
scale factor,%
\[
S_{J}\sim kN_{3}\ln\left[  \frac{\gamma^{3}}{N_{J}}\right]  \text{;}%
\]
via the cubic term; it \emph{favors the coalescence of three} \emph{bispinors
(quarks) into a hadron}.

\item A positive quartic ($N_{4}$) term would create a second basin of
attraction, stabilizing the association of the leptonic and hadronic states
into \emph{Spin}$^{c}$\emph{-4 vortex solitons} [M.C.3]. A negative quartic
term destroys the second basin. The states "slide down the hill", i.e. the
scale factor would expand forever.
\end{enumerate}

\section{Conclusion}

\bigskip

A growing body of evidence suggests that we are immersed in an ocean of "dark
energy" deeper than the matter fields which ride on its surface.

In previous work, we found effective Lagrangians for the standard field
equations- plus the varieties and masses of the particles- by expanding each
of the $\mathbf{8}$ spinor fields as a sum of a global, order $\gamma
^{-\frac{1}{2}}$distribution, plus a local envelope modulation.What happens is
that the envelopes that vary on faster scales are effectively fixed at their
average values, while the solutions that vary on slower scales appear as
"parameters", with slowly- varying values. As the scale factor $\gamma(t)$
increases, the Clifford residues $(\psi^{I}\mathbf{d}\psi_{I}$ $)^{J}$
$\equiv(\mathbf{d\zeta}_{I})^{J}$ become localized over $J$ \ branes, and
their products with the vacuum spinors over $(4-J)$ branes [MC?]. This results
in a\textit{\ spinfoam}, $\Sigma(t).$

If the incidence relations of the $J$ branes in $\Sigma$ are preserved, the
resulting topological charges are \textit{conserved} during the $t$ evolution,
so each brane bounding a volume must have arisen from a small inhomogeneity in
the \textit{initial conditions, }wrapped nontrivially around its boundary,
$\partial\gamma_{5-J}=\gamma_{4-J}.$

In principle, fields in a volume of spacetime may be found by the
\textit{holographic method, }given those on the boundary [refs....]- except
near singularities of the classical spin map, where $\left\vert \mathbf{d\zeta
}\right\vert =0;$ or, near \textit{phase transitions }of the quantum
ensemble.\ Here microscopic inhomogeneities\bigskip\ are "promoted" to the
macroscopic level. Perhaps the galaxies and "great walls" we see today
originated in microscopic vortex lines or planes in the primaeval spinfluid.
But where did\textit{\ these} come from?

Both the classical and quantum cases admit \textit{periodic solutions }for
$N_{4}>0;$ i.e. for matter dominating antimatter. Here, initial conditions at
the "big bang" are inherited from final conditions at the "big crunch".
Neither of these are singular; both are identified with the global turning
point $\gamma_{\min},$ where small perturbations can nucleate large structures
or events. But there are many local turning points-degenerate singularities,
$\left\vert \mathbf{d}^{2}\varsigma\right\vert =0,$ where tiny perturbations-
either random or concious, can change history.

Some conceptual problems arise with the quantum evolution - the sum over
histories in "imaginary time", $T$. The first involves \textit{ergodicity}.
This says that the time average equals the ensemble average; so, to calculate
any function of state, average over all microstates with the same macroscopic
value of the classical observable. For ergodicity to hold literally, the
system must explore \emph{all} \emph{the microstates} compatible with the
observed macrostate during the observation period. Some questions arise about this.

What does ergodicity mean when the "micostates" are different histories of the
universe as a whole? Is the cosmic history we experience actually an average
over all \emph{possible} histories? Does history branch into many worlds at
each decision point; or is one particular path in the ensemble always chosen
(as in a random walk), but the invisible influences that determine
\emph{which} path are so manifold and subtle, and their fluctuations so fast,
that we can only \emph{predict} ensemble averages?

Doesn't Bell's theorem rule out such a "realistic" explanation of which path
is chosen for the "real" history? No; not if its \emph{nonlocal}---i.e. if
there is some globally---determined field; or if there are closed causal
cycles, $\gamma_{1}=\partial B_{2}$, that loop around some area in
\emph{spacetime} to return to the same place, $\mathbf{x,}$\textbf{\ }at the
\emph{same cosmic time,} $T:$

\begin{center}
$\left(  T,\mathbf{x}\right)  \left(  t_{0}+\bigtriangleup t\right)  =\left(
T,\mathbf{x}\right)  \left(  t_{0}\right)  $.
\end{center}

The evidence for such closed \emph{cycles} $\gamma_{1}:t\longrightarrow\left(
T,\mathbf{x}\right)  \left(  t\right)  $ has been staring us in the face for a
century: quantization of action around space\textit{time} cycles:

\begin{center}
\bigskip$\int_{\gamma_{1}}Edt-Pdq=n\pi\hbar,$
\end{center}

and of electric charge, the space\textit{time}\textbf{-}component of the spin
curvature [ccq].

\bigskip Null geodesics, or photons (helieity -1) on $S_{1}\times S_{3},$
propagate once around their twisted light cones, before coming back on their
original values. But \textit{spinors }(helicity-1/2) must go twice around
$\gamma_{1}$ (or once in opposite directions) before coming back on themselves
in value.

These global cycles together with all the local integral-homonymy cycles, make
a dynamical history. In the 4-dimensional picture, history is
\textit{geometry}. Much like Kirchoffs laws for electric circuits, the
physical "laws"are incidence relations between $(4-J)$ branes, each carrying a
dual $J $ form current. Each microhistory must obey these constraints. But the
question remains:

\begin{center}
\textit{\ Why does an observation "collapse" the ensemble of allowed
microstates into a single one?"}
\end{center}

\bigskip We shall see in the sequel that dynamical collapse of the wavepacket
in Minkowshy time, t, can appear instantaneous in cosmic time.

\bigskip

\section{\bigskip bibliography}

\bigskip

\end{document}